\documentclass[a4paper]{article}
\usepackage{hyperref}
\usepackage{multirow}
\usepackage{dblfloatfix}

\usepackage{makecell}
 
\usepackage{cite}
\usepackage[table,xcdraw]{xcolor}
\usepackage{booktabs,subcaption,amsfonts,dcolumn}

\usepackage{INTERSPEECH2019}
\title{WeSinger 2: Fully Parallel Singing Voice Synthesis via Multi-Singer Conditional Adversarial Training}
\name{Zewang Zhang, Yibin Zheng, Xinhui Li, Li Lu}
\address{
  Tencent Inc, China}
\email{\{zewangzhang, hiccupli\}@tencent.com}
\begin{document}

\maketitle
\begin{abstract}
This paper aims to introduce a robust singing voice synthesis (SVS) system to produce very natural and realistic singing voices efficiently by leveraging the adversarial training strategy. On one hand, we designed simple but generic random area conditional discriminators to help supervise the acoustic model, which can effectively avoid the over-smoothed spectrogram prediction and improve the expressiveness of SVS. On the other hand, we subtly combined the spectrogram with the frame-level linearly-interpolated F0 sequence as the input for the neural vocoder, which is then optimized with the help of multiple adversarial conditional discriminators in the waveform domain and multi-scale distance functions in the frequency domain. The experimental results and ablation studies concluded that, compared with our previous auto-regressive work, our new system can produce high-quality singing voices efficiently by fine-tuning different singing datasets covering from several minutes to a few hours. A large number of synthesized songs with different timbres are available online\footnote{https://zzw922cn.github.io/wesinger2} and we highly recommend readers to listen to them.
\end{abstract}

\section{Introduction}
Given the lyrics and musical scores, an essential singing voice synthesis (SVS) system is expected to generate singing voices equipped with accurate pitch, natural rhythm, and high-fidelity waveform. As a consequence of the improvements~\cite{wang2017tacotron,ren2020fastspeech} in the text-to-speech (TTS) field, many researchers employed deep neural networks subsequently for acoustic modeling in SVS systems~\cite{kim2018korean,yi2019singing,lu2020xiaoicesing}, which undoubtedly outperform the traditional SVS methods~\cite{bonada2016expressive,nakamura2014hmm} largely concerning the quality and naturalness. Recently, some neural SVS systems~\cite{gu2021bytesing,zhang2022wesinger} were built by adopting modern acoustic models and auto-regressive neural vocoders~\cite{kalchbrenner2018efficient,valin2019lpcnet}, and they provided good paths to the synthesized singing voices with both higher naturalness and better quality. 
Despite this, some serious drawbacks cannot be ignored in terms of the inefficient synthesis speed and over-smoothing spectrogram prediction~\cite{chen2020hifisinger}.

In recent years, generative adversarial network (GAN)~\cite{goodfellow2014generative} becomes well-known for its superior performance in implicitly capturing the distribution of the observed data with the necessary guidance from an adversarial discriminator. According to this, how to combine GANs with modern model designs to attain more realistic singing synthesis has been widely studied, such as leveraging different variants of GAN with the lyrics and musical scores as conditions for better spectrogram prediction~\cite{hono2019singing,chandna2019wgansing,choi2020korean,wu2020adversarially} or designing effective up-sampling networks with GAN for high-fidelity singing waveform generation~\cite{liu21e_interspeech,chen2021singgan,huang2021multi}. These methods either employed some autoregressive modules or have not applied GAN to both spectrogram prediction and waveform generation simultaneously. More importantly, there are still some obvious flaws in the generated singing voices. Until recently, there have been several SVS systems that apply adversarial training for optimizing both acoustic models and neural vocoders, such as HiFiSinger~\cite{chen2020hifisinger} and N-Singer~\cite{lee2021n}. Both of them shared the same structure in the feed-forward Transformer (FFT) acoustic model and Parallel WaveGAN~\cite{yamamoto2020parallel} vocoder, however, they did not investigate the multi-singer training under the GAN framework. By the way, another GAN-based end-to-end singing synthesis system VISinger~\cite{zhang2022visinger} has been proposed recently to improve the performance of the traditional two-stage method. To our knowledge, a considerable number of synthesized singing samples still sound unnatural.

Inspired by the above studies, in this paper, we propose a fully parallel multi-singer SVS system and employ adversarial training to prompt the realistic spectrogram and waveform generation sequentially. The highlights of our work can be summarized as follows:
\begin{itemize}
  \item We propose an efficient and robust multi-singer SVS system composed of fully parallel architectures without any recurrent unit, and it is feasible to do singer adaptation with limited singing data.
  \item A generic, simple but effective couple of multi random area discriminators with each singer's identity as the condition are leveraged to train the acoustic model jointly with the traditional L1 loss, which can significantly ease the over-smoothing problem and improve the quality of the predicted Mel-spectrogram. By the way, we adopt an effective multi-receptive dilated convolutional Postnet following the FFT-based acoustic model.
  \item With the linearly-interpolated F0 sequence and Mel-spectrogram produced by the acoustic model, a carefully designed GAN-based vocoder is introduced, which works well for capturing the different distributions of many singers' data. To enhance the performance and training stability, we leverage the speaker identity information to the intermediate layer of each discriminator.
\end{itemize}
Our proposed system is an extension work with our previous work WeSinger~\cite{zhang2022wesinger}, which is a data-augmented SVS system with pitch-weighted progressive loss and takes the 24 kHz LPCNet as the neural vocoder. This work shares many hyper-parameter settings, data augmentation techniques, and auxiliary loss functions with WeSinger~\cite{zhang2022wesinger}. Therefore, our proposed system in this paper is named WeSinger 2. Experimental results indicated that WeSinger 2 can generate natural and high-quality singing voices more efficiently, and several ablation studies demonstrated the effectiveness of our designs.

\section{Methodology}
\label{section2}
\subsection{Architecture Overview}
\begin{figure}[htp]
    \centering
    \includegraphics[width=8cm]{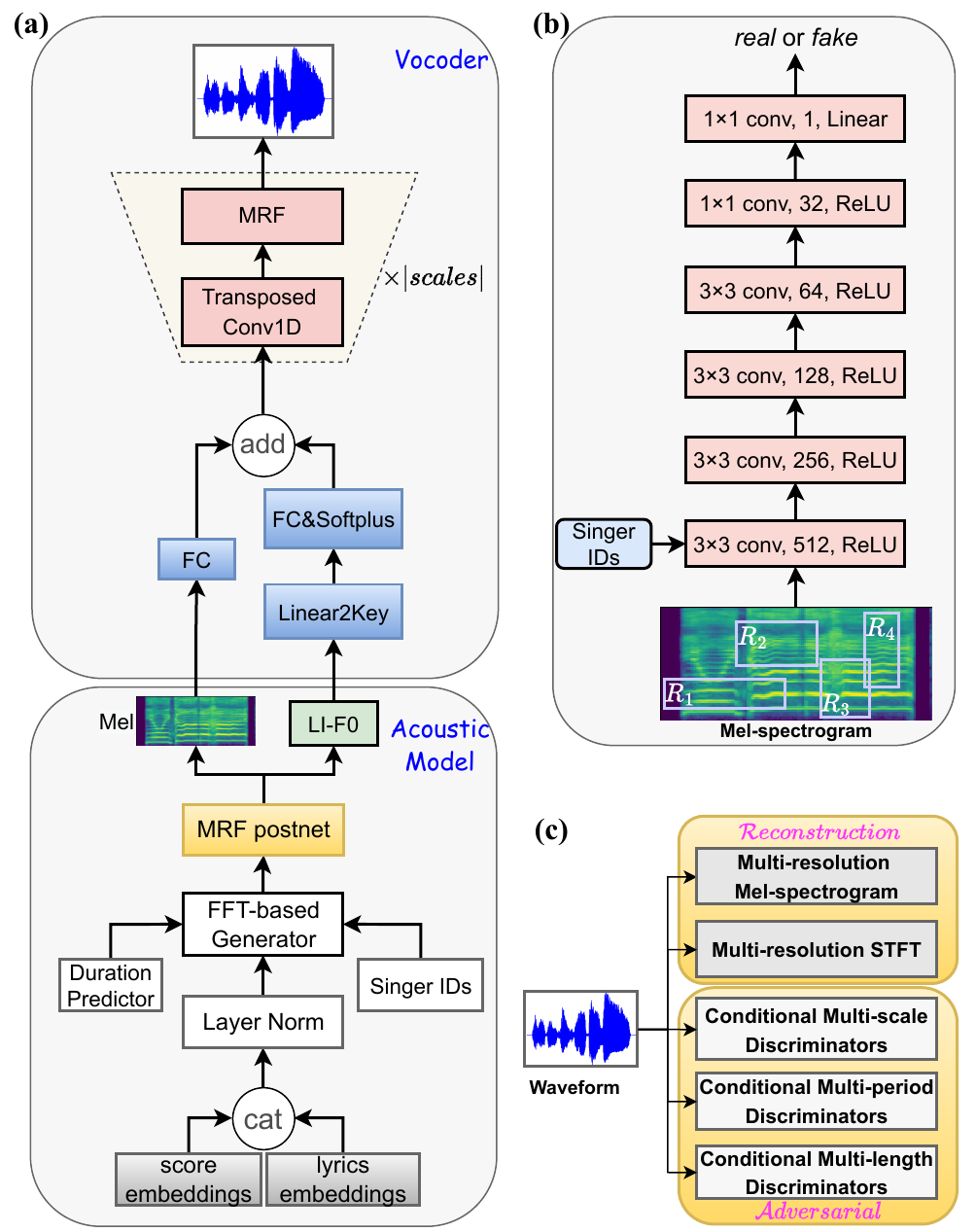}
    \caption{\textbf{(a)} The overall architecture of our proposed multi-singer SVS system WeSinger 2 at the inference stage. \textbf{(b)} The conditional multi random area discriminators (MRADs) designed for training the acoustic model. \textbf{(c)} The reconstruction and adversarial functions adopted for training the GAN-based neural vocoder. The gradient reversal layer for singer classification during multi-singer training and the progressive decoder loss are omitted here, which have been described in~\cite{zhang2022wesinger}.}
    \label{fig:wesinger2}
\end{figure}

The general architecture of WeSinger 2 is illustrated in Figure~\ref{fig:wesinger2}(a). The input representation and the FFT-based encoder follow the designs of WeSinger~\cite{zhang2022wesinger}. Instead of adopting a BLSTM-based duration predictor as WeSinger, here we replace the LSTM layers with several 1-D convolutional (Conv1D) layers inspired by~\cite{ren2020fastspeech} for faster inference speed and similar performance. The Mel-spectrogram with linearly-interpolated F0 (denoted as ``LI-F0'' in Figure~\ref{fig:wesinger2}(a)) sequence is produced from FFT blocks and improved by an effective Post-net based on multi-receptive field (MRF) instead of the CBHG Post-net. In addition, the LPCNet vocoder adopted by WeSinger~\cite{zhang2022wesinger} is also replaced with a parallel GAN-based vocoder to efficiently produce high-fidelity waveforms. We will introduce the key improvements of WeSinger 2 in the following subsections.
\subsection{Acoustic Model}
Based on WeSinger~\cite{zhang2022wesinger}, we made two major improvements to the FFT-based acoustic model as followings, including the MRF-based Post-net and the conditional adversarial training.
\begin{figure*}[htp]
\begin{subfigure}{.33\textwidth}
  \centering
  \includegraphics[width=.8\linewidth]{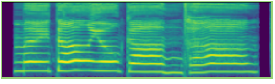}  
  \caption{ground-truth}
  \label{fig:sub-first}
\end{subfigure}
\begin{subfigure}{.33\textwidth}
  \centering
  \includegraphics[width=.8\linewidth]{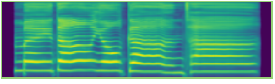}  
  \caption{L1 loss}
  \label{fig:sub-second}
\end{subfigure}
\begin{subfigure}{.33\textwidth}
  \centering
  \includegraphics[width=.8\linewidth]{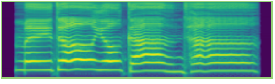}  
  \caption{L1 + adv. loss}
  \label{fig:sub-second}
\end{subfigure}
\caption{The quality improvement in the predicted Mel-spectrogram by WeSinger 2 with multi random area discriminators (MRADs).}
\label{fig:mel_spectrogram}
\end{figure*}

\textit{(1) \textbf{MRF-based Post-net}} It is an important and necessary design to append Post-net to the acoustic model for improving the quality of predicted Mel-spectrogram~\cite{shen2018natural}. Inspired by~\cite{kong2020hifi}, we design a deep residual Conv1D-based Post-net based on multi-receptive fields to enhance the performance of contextual modeling. Specifically, the Post-net in WeSinger 2 is composed of three blocks with different convolutional kernel sizes, each block is composed of three Conv1D layers with different dilation rates. Each Conv1D layer is followed by leaky rectified linear unit (ReLU) activation. To ease the optimization, we add residual connections between all adjacent Conv1D layers and blocks. Compared with the deep CNNs-based Post-net adopted in~\cite{lee2021n,shen2018natural}, our MRF-based Post-net can be more effective in reproducing Mel-spectrogram with around the same computational cost.

\textit{(2) \textbf{Conditional Adversarial Training}} We found that optimizing the duration-allocated acoustic model by only L1 construction loss could lead to fuzzy and over-smoothing Mel-spectrogram prediction, which will either harm the coherence of waveforms generated from the neural vocoder or make the synthesized singing voices lack expressiveness, such as trill and glissando in singing. To alleviate these defects in singing synthesis, we combine adversarial training with L1 reconstruction loss to help train the acoustic model. Different from the sub-frequency discriminator~\cite{chen2020hifisinger} and voice-aware conditional discriminator~\cite{lee2021n}, we design the conditional
multi random area discriminators (MRADs) to effectively discriminate different structures in the Mel-spectrogram of each singer, which is also more generic and easier to implement. As shown in Figure~\ref{fig:wesinger2}(b), each discriminator, composed of six two-dimensional convolutional (Conv2D) layers, takes as input a random rectangle region from the Mel-spectrogram and predicts a scalar value to indicate whether the selected region is real or fake. The height of each randomly selected region represents the frequency band and the width represents the number of frames. We empirically fix the values of height to be $[20, 30, 40, 50]$ and width to be $[190, 160, 70, 30]$. The intuition behind this design is to allow MRADs effectively capture features at different resolutions in both the time domain and frequency domain. The location of each rectangle region is randomly sampled at every training step to achieve the effect of data augmentation. Besides, we found that adding the singer embedding vectors to the output of the first Conv2D layer can both 
prevent the generator from overfitting to a specified singer's data and improve the quality of the predicted Mel-spectrogram. For adversarial training, we use the popular objective functions proposed in LSGAN~\cite{mao2017least} and combine them with the L1 reconstruction loss as follows:
\begin{equation}
\begin{split}
 \mathcal{L}_{block}(D;c) &=\mathbb{E}_{y\sim p_{data}(y)}[\sum_{i=1}^{4}(D_i(y, c)-1)^2] \\
&+\mathbb{E}_{x\sim p_{g}(x)}[\sum_{i=1}^{4}(D_i(x, c))^2] \\
\mathcal{L}_{block}(G;c) &=\lambda_{adv}\mathbb{E}_{x\sim p_{g}(x)}[\sum_{i=1}^{4}(D_i(x, c)-1)^2]+\lambda_{l}|x - y| \\
\end{split}
\end{equation}
where $D_i$ denotes the $i$-th discriminator, $G$ represents the FFT-based acoustic model, $c$ denotes the identity of each singer, $x$ is the generated Mel-spectrogram from each decoder block, $y$ is the ground-truth Mel-spectrogram, $p_{data}$ is the distribution of ground-truth Mel-spectrogram, $p_{g}$ is the distribution of generated Mel-spectrogram from each decoder block, $|x-y|$ is the L1 reconstruction loss between the predicted Mel-spectrogram from each decoder block and the ground-truth Mel-spectrogram, and $\mathcal{V}_{block}$ indicates the objective function of each decoder block. $\lambda_{adv}$ and $\lambda_{l}$ are the coefficients for adversarial loss and L1 reconstruction loss, respectively. We set $\lambda_{l}=20$ in this paper. $\lambda_{adv}$ is set to be $1$ when pre-training and to be $0.7$ when fine-tuning. We also adopt the progressive decoder loss here, the effectiveness of which has been demonstrated in ~\cite{zhang2022wesinger}. Therefore, the overall objective functions for the acoustic model's generator $\mathcal{V}_{am}(G)$ and discriminators $\mathcal{V}_{am}(D)$ can be formulated as:
\begin{equation}
\begin{split}
 \mathcal{V}_{am}(D) &=\mathop{min}\limits_{D} \sum_{block=1}^{B} \mathcal{L}_{block}(D;c) \\
 \mathcal{V}_{am}(G) &=\mathop{min}\limits_{G} \sum_{block=1}^{B} \mathcal{L}_{block}(G;c) \\
\end{split}
\end{equation}
where $B$ denotes the number of all decoder blocks and Post-net. Similar as~\cite{zhang2022wesinger}, the duration loss and the adversarial singer classifier loss are also adopted for training and we omit both terms here.

\subsection{Neural Vocoder}
\textit{(1) \textbf{Waveform Generator}} Inspired by HiFiGAN~\cite{kong2020hifi}, we design a new GAN-based neural vocoder for SVS as shown in Figure~\ref{fig:wesinger2}(a). Considering that singing voices usually have a wide range of pitch and longer vowel duration~\cite{lee2021n} than speech, we made several improvements in the design of the generator compared with the original HiFiGAN. First, in addition to the commonly-used Mel-spectrogram, our vocoder also considers as input the 88 discrete piano keys which are quantized from the linearly-interpolated F0 sequence according to the MIDI standard~\cite{midi1996complete}. The discrete keys are then embedded by looking up a trainable matrix $F\in \mathbb{R}^{88\times d}$, in which $d$ denotes the size of the embedding vector. We set $d=16$ empirically in this paper and observed that a larger value of $d$ 
could harm the performance. Second, to avoid the instability of different singers' corpus during training, we employ a fully-connected layer with \textit{softplus} activation function for the key embedding sequence, and the Mel-spectrogram is also transformed by a linear fully-connected layer, then we take the summation of them as input to the following convolution layers. We found that such designs can both stabilize training and improve the quality of generated singing 
waveforms compared with concatenation or summation directly. 

\textit{(2) \textbf{Discriminative Training}} As for optimizing the generator (shown in Figure~\ref{fig:wesinger2}(c)), we adopt the multi-scale discriminator (MSD)~\cite{kumar2019melgan} to capture long-term dependencies and multi-period discriminator (MPD)~\cite{kong2020hifi} to handle the diverse periodic patterns in the audio signal. Besides, we also introduce the multi-length discriminator (MLD) proposed in~\cite{chen2020hifisinger} by taking audio waveform as input of several Conv1D layers with linearly increasing dilation rates. Different from previous works, here we inject the singer identity information to the specified intermediate layer of each discriminator, which can utilize the multi-singer pre-training effectively and achieve training the generator with larger kernel size and dilation rate stably. In addition to the conditional discriminators, we apply both the multi-resolution STFT~\cite{yang2021multi} and multi-resolution Mel-spectrogram~\cite{kong2020hifi} reconstruction loss functions to help train the generator more quickly. Overall, the final training objectives for the generator $\mathcal{V}_{voc}(G)$ and the discriminators $\mathcal{V}_{voc}(D)$ can be formulated as: 
\begin{equation}
\begin{split}
\mathcal{L}_{adv}(D;c) &= \mathbb{E}_{y\sim p_{data}(y)}[(D(y, c)-1)^2] \\
&+ \mathbb{E}_{x\sim p_{g}(x)}[(D(x, c))^2],\forall D\in\mathcal{S} \\
\mathcal{L}_{adv}(G;c) &= \mathbb{E}_{x\sim p_{g}(x)}[\sum_{D\in\mathcal{S}}(D(x, c)-1)^2]\\
\mathcal{L}_{recons}(G) &= \lambda_{s}\mathcal{L}_{stft} + \lambda_{m}\mathcal{L}_{mel} \\
 \mathcal{V}_{voc}(D) &= \mathop{min}\limits_{D} \mathcal{L}_{adv}(D;c) \\
 \mathcal{V}_{voc}(G) &= \mathop{min}\limits_{G}[ \mathcal{L}_{adv}(G;c) + \mathcal{L}_{recons}(G)] \\
\end{split}
\end{equation}
in which $\mathcal{S}$=\{MSD, MPD, MLD\}, $p_{data}$ is the distribution of the ground-truth waveform, $p_{g}$ is the distribution of the generated waveform. $\lambda_{s}$ and $\lambda_{m}$ are the scale factors for both the multi-resolution STFT reconstruction loss and the multi-resolution Mel-spectrogram reconstruction loss, respectively. We obtained the optimal performance of the proposed vocoder by setting $\lambda_{s}=\lambda_{m}=15$ in this paper. 
\section{Experiments}
\subsection{Datasets}
\label{datasets}
We follow our previous work WeSinger~\cite{zhang2022wesinger} to use a 30-hour multi-singer singing corpus for pre-training WeSinger 2. With the pre-trained model, we try to fine-tune it on three processed datasets (24 kHz with 16-bit quantization): (1) \textit{\textbf{Opencpop}}~\cite{wang2022opencpop}, a public female singing dataset\footnote{https://wenet.org.cn/opencpop/} including 3550 segments for training and 206 segments for testing; (2) \textit{\textbf{Data-L}}, a large-scale high-quality singing dataset including 4700 segments for training and 298 segments for testing; (3) \textit{\textbf{Data-S}}, a small-scale low-quality amateur singing dataset including 348 segments for training and 10 segments for testing. To improve the diversity, each recording (except the recordings from \textit{Opencpop}) is split into audio segments according to the \textit{VS}-augmented~\cite{zhang2022wesinger} strategy. Mel-spectrograms with 80 bands are extracted as the intermediate feature representation. By the way, the F0 contours were extracted with the YIN~\cite{de2002yin} algorithm and we replaced the unvoiced parts with the linearly-interpolated F0 values. Notably, both Mel-spectrograms and F0 values are rescaled with the mean-variance normalization trick individually.

\subsection{Singer Adaptation Setup}
We compared WeSinger 2 with our previous work WeSinger~\cite{zhang2022wesinger}. All acoustic models and neural vocoders were first pre-trained for 1200k steps with a batch size of 24 on the 30-hour multi-singer corpus. The initial learning rate was set to 8e-4 with a warm-up strategy at the beginning of 150k steps and then decayed until the final learning rate of 1e-4 for the following 1050k steps. The pre-trained models were then fine-tuned with a batch size of 24 and a fixed learning rate of 1e-4 on three datasets as described in Section~\ref{datasets}. 
To make a trade-off between the generalization and high fidelity, we simplified the sampling strategy proposed in~\cite{Zheng2022Zero} by selecting a sample of the target singer with a probability of $0.7$ and a sample of other singers with a probability of $0.3$ to prepare the training dataset for fine-tuning. With such a simple sampling strategy during fine-tuning, it is feasible to build a robust SVS system without obvious harm to the timbre of the target singer especially when the training data of the target singer is very limited. We leave the detailed investigation of efficient few-shot SVS for future work.

\subsection{Evaluations}
All short sentences were generated from the fine-tuned systems with the ground-truth duration and then concatenated into singing segments. By the way, the performance of singing synthesis with predicted duration has been shown on our demo page\footnote{https://zzw922cn.github.io/wesinger2}. The evaluation for the performance of both WeSinger~\cite{zhang2022wesinger} and WeSinger 2 is two folds. Subjectively, we conducted the Mean Opinion Score (MOS) test, in which twenty listeners were asked to rate each audio segment on a scale from 1 to 5 in terms of overall quality. Objectively, we calculate F0 RMSE (F0 root mean square error) and MSD (Mel-spectrogram distortion) between the
generated voices and recordings to distinguish whether the generated voice is out-of-tune or has poor quality.
\begin{table}[]
\centering
\arrayrulewidth1.0pt
\caption{Quantitative and qualitative evaluation results of both WeSinger and WeSinger 2 on three different datasets.}
\begin{tabular}{ccccc}
corpus & \multicolumn{1}{c}{system} & MOS & \begin{tabular}[c]{@{}c@{}}F0 \\ RMSE\end{tabular} & MSD \\ \hline
& GT & $3.91\pm0.10$ & $-$ & $-$ \\
 & WeSinger & $3.21\pm0.15$ & $17.51$ & $1.40$ \\
\multirow{-3}{*}{ \textit{Opencpop}} & \cellcolor[HTML]{EFEFFF}\textbf{WeSinger 2} & \cellcolor[HTML]{EFEFFF} $\textbf{3.40}\pm\textbf{0.06}$ & \cellcolor[HTML]{EFEFFF}$\textbf{16.61}$ & \cellcolor[HTML]{EFEFFF}$\textbf{1.31}$ \\ \hline
& GT & $4.10\pm0.06$ & $-$ & $-$ \\
 & WeSinger & $3.60\pm0.14$ & $15.16$ & $1.53$ \\ 
\multirow{-3}{*}{\textit{Data-L}} & \cellcolor[HTML]{EFEFFF}\textbf{WeSinger 2} & \cellcolor[HTML]{EFEFFF}$\textbf{3.71}\pm\textbf{0.07}$ & \cellcolor[HTML]{EFEFFF}$\textbf{14.68}$ & \cellcolor[HTML]{EFEFFF}$\textbf{1.41}$ \\ \hline
& GT & $3.75\pm0.15$ & $-$ & $-$ \\
 & WeSinger & $3.21\pm0.15$ & $13.23$ & $2.64$ \\ 
\multirow{-2}{*}{\textit{Data-S}} & \cellcolor[HTML]{EFEFFF}\textbf{WeSinger 2} & \cellcolor[HTML]{EFEFFF}$3.19\pm0.04$ & \cellcolor[HTML]{EFEFFF}$\textbf{12.32}$ & \cellcolor[HTML]{EFEFFF}$\textbf{1.32}$ \\ 
\end{tabular}
\label{MOS_for_each_singer}
\begin{flushleft}
\footnotesize{\textbf{NOTE}: The duration accuracy is not shown here as the duration model behaves on par with that in~\cite{zhang2022wesinger}. The singing voices generated with predicted duration will be also available on our demo page.}\\
\end{flushleft}
\end{table}
\begin{table}[]
\centering
\arrayrulewidth1.0pt
\caption{Parameters and efficiency of WeSinger and WeSinger 2 at the synthesis stage.}
\begin{tabular}{cccc}
system & \begin{tabular}[c]{@{}c@{}}\#param \\ \textit{A.M.}\textsuperscript{\textdagger} \end{tabular} & \begin{tabular}[c]{@{}c@{}}\#param \\ \textit{Voc.}\textsuperscript{\textdaggerdbl} \end{tabular} & \begin{tabular}[c]{@{}c@{}}Speed on CPU\textsuperscript{\textsection}\\(kHz/s) \end{tabular} \\ \hline
WeSinger &37 M&2.49 M& 4.01\\
\cellcolor[HTML]{EFEFFF}\textbf{WeSinger 2} & \cellcolor[HTML]{EFEFFF}\textbf{49 M} & \cellcolor[HTML]{EFEFFF}\textbf{5.89 M} & \cellcolor[HTML]{EFEFFF}\textbf{65.24} \\ 
\end{tabular}
\label{systems}
\begin{flushleft}
\footnotesize{\textsuperscript{\textdagger} Acoustic model. \textsuperscript{\textdaggerdbl} Neural vocoder. \textsuperscript{\textsection} Tested on a single core of AMD EPYC 7K62 CPU @ 2.60 GHz in Python Environment. Notably, all discriminators are not considered here.}\\
\end{flushleft}
\end{table}
As shown in Table~\ref{MOS_for_each_singer}, compared with WeSinger~\cite{zhang2022wesinger}, WeSinger 2 has the higher MOS with significantly higher confidence on both \textit{Opencpop} and \textit{Data-L} datasets, and around the same MOS on the \textit{Data-S} dataset. Most listeners made comments that the key improvement with WeSinger 2 is that it maintains a very stable performance at musical notes with long vowels, which can also be indicated by the lower MSD value. In addition, WeSinger 2 behaves much better than WeSinger~\cite{zhang2022wesinger} in terms of some rare and hard musical notes, which can be reflected in the lower F0 RMSE value. Besides, we found that most of the generated singing audios by WeSinger 2 do not have metallic or inconsistent artifacts as found in~\cite{wang2022opencpop} and do not sound unnatural. Apart from the performance, 
around 3 times faster than real-time synthesis can be achieved on a single core of a moderate CPU @ 2.60 GHz in a Python environment. The number of parameters and synthesis speed of both WeSinger and WeSinger 2 are listed in Table~\ref{systems}.
\begin{table}[!tb]
\arrayrulewidth1.0pt
\begin{minipage}[b]{0.45\linewidth}\centering
\caption{Ablation study for the acoustic model of WeSinger 2.}
\begin{tabular}{lc}
\multicolumn{1}{c}{case} & CMOS\textdownarrow \\ \hline
\rowcolor[HTML]{EFEFFF}\textbf{proposed} & 0 \\
w/o MRF\textsuperscript{\textdagger} & $-0.10$ \\
w/o condition & $-0.15$ \\
w/o adv. loss & $-0.27$
\end{tabular}
\label{abalation_1}
\end{minipage}
\hspace{0.5cm}
\begin{minipage}[b]{0.45\linewidth}\centering
\caption{Ablation study for the neural vocoder of WeSinger 2.}
\begin{tabular}{lc}
\multicolumn{1}{c}{case} & CMOS\textdownarrow \\ \hline
\rowcolor[HTML]{EFEFFF}\textbf{proposed} & 0 \\
w/o condition & $-0.10$ \\
w/o WLD & $-0.14$ \\
w/o pitch & $-0.16$
\end{tabular}
\label{abalation_2}
\end{minipage}
\begin{flushleft}
\footnotesize{\textsuperscript{\textdagger} It means adopting deep CNNs as Post-net for the acoustic model.}\\
\end{flushleft}
\end{table}
\subsection{Ablation Studies}
\label{ablation}
To verify the effectiveness of our designs in WeSinger 2, we further conduct several ablation studies for the acoustic model and neural vocoder, respectively. We do training on the \textit{Data-L} dataset and choose the Comparative Mean Opinion Score (CMOS) as the evaluation metric. As shown in Table~\ref{abalation_1}, that conditional adversarial training with MRADs plays an important role in better performance, which can alleviate the over-smoothing problem and produce more realistic Mel-spectrograms as illustrated in Figure~\ref{fig:mel_spectrogram}. Meanwhile, replacing the deep CNNs with MRF Post-net can also 
result in a gain of $0.1$ CMOS without additional computational cost. As for the GAN-based neural vocoder (shown in Table~\ref{abalation_2}), we found that removing the singer's identity for discriminators can decrease the high-fidelity of generated audios, and removing the MLD and the key embedding part can also lead to a decrease of around 0.16 CMOS.
\section{Conclusion}
In this paper, we introduced a new SVS system named WeSinger 2 to produce high-quality singing voices efficiently. We proposed the simple but generic multi random area discriminators to achieve the realistic Mel-spectrogram generation and described how to convert the Mel-spectrogram with a frame-level linearly-interpolated F0 sequence into singing waveforms with the proposed GAN-based neural vocoder. Evaluation results indicate that WeSinger 2 can synthesize natural singing voices. The robustness and efficiency of WeSinger 2 make it easy to be deployed in the production environment.

\vfill\pagebreak

\clearpage 
\bibliographystyle{IEEEtran}

\bibliography{mybib}

\begin{thebibliography}{10}
\providecommand{\url}[1]{#1}
\csname url@samestyle\endcsname
\providecommand{\newblock}{\relax}
\providecommand{\bibinfo}[2]{#2}
\providecommand{\BIBentrySTDinterwordspacing}{\spaceskip=0pt\relax}
\providecommand{\BIBentryALTinterwordstretchfactor}{4}
\providecommand{\BIBentryALTinterwordspacing}{\spaceskip=\fontdimen2\font plus
\BIBentryALTinterwordstretchfactor\fontdimen3\font minus
  \fontdimen4\font\relax}
\providecommand{\BIBforeignlanguage}[2]{{%
\expandafter\ifx\csname l@#1\endcsname\relax
\typeout{** WARNING: IEEEtran.bst: No hyphenation pattern has been}%
\typeout{** loaded for the language `#1'. Using the pattern for}%
\typeout{** the default language instead.}%
\else
\language=\csname l@#1\endcsname
\fi
#2}}
\providecommand{\BIBdecl}{\relax}
\BIBdecl

\bibitem{wang2017tacotron}
Y.~Wang, R.~Skerry-Ryan, D.~Stanton, Y.~Wu, R.~J. Weiss, N.~Jaitly, Z.~Yang,
  Y.~Xiao, Z.~Chen, S.~Bengio, Q.~Le, Y.~Agiomyrgiannakis, R.~Clark, and R.~A.
  Saurous, ``{Tacotron: Towards End-to-End Speech Synthesis},'' in \emph{Proc.
  Interspeech 2017}, 2017, pp. 4006--4010.

\bibitem{ren2020fastspeech}
Y.~Ren, C.~Hu, X.~Tan, T.~Qin, S.~Zhao, Z.~Zhao, and T.-Y. Liu, ``Fastspeech 2:
  Fast and high-quality end-to-end text to speech,'' \emph{arXiv preprint
  arXiv:2006.04558}, 2020.

\bibitem{kim2018korean}
J.~Kim, H.~Choi, J.~Park, S.~Kim, J.~Kim, and M.~Hahn, ``Korean singing voice
  synthesis system based on an lstm recurrent neural network,'' in \emph{Proc.
  Interspeech}, 2018, pp. 1551--1555.

\bibitem{yi2019singing}
Y.-H. Yi, Y.~Ai, Z.-H. Ling, and L.-R. Dai, ``{Singing Voice Synthesis Using
  Deep Autoregressive Neural Networks for Acoustic Modeling},'' in \emph{Proc.
  Interspeech 2019}, 2019, pp. 2593--2597.

\bibitem{lu2020xiaoicesing}
P.~Lu, J.~Wu, J.~Luan, X.~Tan, and L.~Zhou, ``{XiaoiceSing: A High-Quality and
  Integrated Singing Voice Synthesis System},'' in \emph{Proc. Interspeech
  2020}, 2020, pp. 1306--1310.

\bibitem{bonada2016expressive}
J.~Bonada, M.~Umbert, and M.~Blaauw, ``{Expressive Singing Synthesis Based on
  Unit Selection for the Singing Synthesis Challenge 2016},'' in \emph{Proc.
  Interspeech 2016}, 2016, pp. 1230--1234.

\bibitem{nakamura2014hmm}
K.~Nakamura, K.~Oura, Y.~Nankaku, and K.~Tokuda, ``Hmm-based singing voice
  synthesis and its application to japanese and english,'' in \emph{2014 IEEE
  International Conference on Acoustics, Speech and Signal Processing
  (ICASSP)}.\hskip 1em plus 0.5em minus 0.4em\relax IEEE, 2014, pp. 265--269.

\bibitem{gu2021bytesing}
Y.~Gu, X.~Yin, Y.~Rao, Y.~Wan, B.~Tang, Y.~Zhang, J.~Chen, Y.~Wang, and Z.~Ma,
  ``Bytesing: A chinese singing voice synthesis system using duration allocated
  encoder-decoder acoustic models and wavernn vocoders,'' in \emph{2021 12th
  International Symposium on Chinese Spoken Language Processing
  (ISCSLP)}.\hskip 1em plus 0.5em minus 0.4em\relax IEEE, 2021, pp. 1--5.

\bibitem{zhang2022wesinger}
Z.~Zhang, Y.~Zheng, X.~Li, and L.~Lu, ``{WeSinger: Data-augmented Singing Voice
  Synthesis with Auxiliary Losses},'' in \emph{Proc. Interspeech 2022}, 2022,
  pp. 4252--4256.

\bibitem{kalchbrenner2018efficient}
N.~Kalchbrenner, E.~Elsen, K.~Simonyan, S.~Noury, N.~Casagrande, E.~Lockhart,
  F.~Stimberg, A.~Oord, S.~Dieleman, and K.~Kavukcuoglu, ``Efficient neural
  audio synthesis,'' in \emph{International Conference on Machine
  Learning}.\hskip 1em plus 0.5em minus 0.4em\relax PMLR, 2018, pp. 2410--2419.

\bibitem{valin2019lpcnet}
J.-M. Valin and J.~Skoglund, ``Lpcnet: Improving neural speech synthesis
  through linear prediction,'' in \emph{ICASSP 2019-2019 IEEE International
  Conference on Acoustics, Speech and Signal Processing (ICASSP)}.\hskip 1em
  plus 0.5em minus 0.4em\relax IEEE, 2019, pp. 5891--5895.

\bibitem{chen2020hifisinger}
J.~Chen, X.~Tan, J.~Luan, T.~Qin, and T.-Y. Liu, ``Hifisinger: Towards
  high-fidelity neural singing voice synthesis,'' \emph{arXiv preprint
  arXiv:2009.01776}, 2020.

\bibitem{goodfellow2014generative}
I.~Goodfellow, J.~Pouget-Abadie, M.~Mirza, B.~Xu, D.~Warde-Farley, S.~Ozair,
  A.~Courville, and Y.~Bengio, ``Generative adversarial nets,'' \emph{Advances
  in neural information processing systems}, vol.~27, 2014.

\bibitem{hono2019singing}
Y.~Hono, K.~Hashimoto, K.~Oura, Y.~Nankaku, and K.~Tokuda, ``Singing voice
  synthesis based on generative adversarial networks,'' in \emph{ICASSP
  2019-2019 IEEE International Conference on Acoustics, Speech and Signal
  Processing (ICASSP)}.\hskip 1em plus 0.5em minus 0.4em\relax IEEE, 2019, pp.
  6955--6959.

\bibitem{chandna2019wgansing}
P.~Chandna, M.~Blaauw, J.~Bonada, and E.~G{\'o}mez, ``Wgansing: A multi-voice
  singing voice synthesizer based on the wasserstein-gan,'' in \emph{2019 27th
  European Signal Processing Conference (EUSIPCO)}.\hskip 1em plus 0.5em minus
  0.4em\relax IEEE, 2019, pp. 1--5.

\bibitem{choi2020korean}
S.~Choi, W.~Kim, S.~Park, S.~Yong, and J.~Nam, ``Korean singing voice synthesis
  based on auto-regressive boundary equilibrium gan,'' in \emph{ICASSP
  2020-2020 IEEE International Conference on Acoustics, Speech and Signal
  Processing (ICASSP)}.\hskip 1em plus 0.5em minus 0.4em\relax IEEE, 2020, pp.
  7234--7238.

\bibitem{wu2020adversarially}
J.~Wu and J.~Luan, ``{Adversarially Trained Multi-Singer Sequence-to-Sequence
  Singing Synthesizer},'' in \emph{Proc. Interspeech 2020}, 2020, pp.
  1296--1300.

\bibitem{liu21e_interspeech}
Z.~Liu, C.~Miao, Q.~Zhu, M.~Chen, J.~Ma, S.~Wang, and J.~Xiao,
  ``{EfficientSing: A Chinese Singing Voice Synthesis System Using
  Duration-Free Acoustic Model and HiFi-GAN Vocoder},'' in \emph{Proc.
  Interspeech 2021}, 2021, pp. 1609--1613.

\bibitem{chen2021singgan}
F.~Chen, R.~Huang, C.~Cui, Y.~Ren, J.~Liu, and Z.~Zhao, ``Singgan: Generative
  adversarial network for high-fidelity singing voice generation,'' \emph{arXiv
  preprint arXiv:2110.07468}, 2021.

\bibitem{huang2021multi}
R.~Huang, F.~Chen, Y.~Ren, J.~Liu, C.~Cui, and Z.~Zhao, ``Multi-singer: Fast
  multi-singer singing voice vocoder with a large-scale corpus,'' in
  \emph{Proceedings of the 29th ACM International Conference on Multimedia},
  2021, pp. 3945--3954.

\bibitem{lee2021n}
G.-H. Lee, T.-W. Kim, H.~Bae, M.-J. Lee, Y.-I. Kim, and H.-Y. Cho, ``{N-Singer:
  A Non-Autoregressive Korean Singing Voice Synthesis System for Pronunciation
  Enhancement},'' in \emph{Proc. Interspeech 2021}, 2021, pp. 1589--1593.

\bibitem{yamamoto2020parallel}
R.~Yamamoto, E.~Song, and J.-M. Kim, ``Parallel wavegan: A fast waveform
  generation model based on generative adversarial networks with
  multi-resolution spectrogram,'' in \emph{ICASSP 2020-2020 IEEE International
  Conference on Acoustics, Speech and Signal Processing (ICASSP)}.\hskip 1em
  plus 0.5em minus 0.4em\relax IEEE, 2020, pp. 6199--6203.

\bibitem{zhang2022visinger}
Y.~Zhang, J.~Cong, H.~Xue, L.~Xie, P.~Zhu, and M.~Bi, ``Visinger: Variational
  inference with adversarial learning for end-to-end singing voice synthesis,''
  in \emph{ICASSP 2022-2022 IEEE International Conference on Acoustics, Speech
  and Signal Processing (ICASSP)}.\hskip 1em plus 0.5em minus 0.4em\relax IEEE,
  2022, pp. 7237--7241.

\bibitem{shen2018natural}
J.~Shen, R.~Pang, R.~J. Weiss, M.~Schuster, N.~Jaitly, Z.~Yang, Z.~Chen,
  Y.~Zhang, Y.~Wang, R.~Skerrv-Ryan \emph{et~al.}, ``Natural tts synthesis by
  conditioning wavenet on mel spectrogram predictions,'' in \emph{2018 IEEE
  International Conference on Acoustics, Speech and Signal Processing
  (ICASSP)}.\hskip 1em plus 0.5em minus 0.4em\relax IEEE, 2018, pp. 4779--4783.

\bibitem{kong2020hifi}
J.~Kong, J.~Kim, and J.~Bae, ``Hifi-gan: Generative adversarial networks for
  efficient and high fidelity speech synthesis,'' \emph{Advances in Neural
  Information Processing Systems}, vol.~33, pp. 17\,022--17\,033, 2020.

\bibitem{mao2017least}
X.~Mao, Q.~Li, H.~Xie, R.~Y. Lau, Z.~Wang, and S.~Paul~Smolley, ``Least squares
  generative adversarial networks,'' in \emph{Proceedings of the IEEE
  international conference on computer vision}, 2017, pp. 2794--2802.

\bibitem{midi1996complete}
M.~M. Association \emph{et~al.}, ``The complete midi 1.0 detailed
  specification,'' \emph{Los Angeles, CA, The MIDI Manufacturers Association},
  1996.

\bibitem{kumar2019melgan}
K.~Kumar, R.~Kumar, T.~de~Boissiere, L.~Gestin, W.~Z. Teoh, J.~Sotelo,
  A.~de~Br{\'e}bisson, Y.~Bengio, and A.~C. Courville, ``Melgan: Generative
  adversarial networks for conditional waveform synthesis,'' \emph{Advances in
  neural information processing systems}, vol.~32, 2019.

\bibitem{yang2021multi}
G.~Yang, S.~Yang, K.~Liu, P.~Fang, W.~Chen, and L.~Xie, ``Multi-band melgan:
  Faster waveform generation for high-quality text-to-speech,'' in \emph{2021
  IEEE Spoken Language Technology Workshop (SLT)}.\hskip 1em plus 0.5em minus
  0.4em\relax IEEE, 2021, pp. 492--498.

\bibitem{wang2022opencpop}
Y.~Wang, X.~Wang, P.~Zhu, J.~Wu, H.~Li, H.~Xue, Y.~Zhang, L.~Xie, and M.~Bi,
  ``Opencpop: A high-quality open source chinese popular song corpus for
  singing voice synthesis,'' \emph{arXiv preprint arXiv:2201.07429}, 2022.

\bibitem{de2002yin}
A.~De~Cheveign{\'e} and H.~Kawahara, ``Yin, a fundamental frequency estimator
  for speech and music,'' \emph{The Journal of the Acoustical Society of
  America}, vol. 111, no.~4, pp. 1917--1930, 2002.

\bibitem{Zheng2022Zero}
Y.~Zheng, Z.~Zhang, X.~Li, W.~Su, and L.~Lu, ``Zero-shot cross-lingual transfer
  using multi-stream encoder and efficient speaker representation,'' in
  \emph{ICASSP 2022-2022 IEEE International Conference on Acoustics, Speech and
  Signal Processing (ICASSP)}.\hskip 1em plus 0.5em minus 0.4em\relax IEEE,
  2022, pp. 8027--8031.

\end{thebibliography}

\end{document}